\documentclass[11pt]{article}
\usepackage[x11names,table]{xcolor}
\usepackage{coling}
\usepackage{times}
\usepackage{latexsym}
\usepackage[T1]{fontenc}
\usepackage[utf8]{inputenc}
\usepackage{microtype}
\usepackage{inconsolata}

\usepackage{xspace}
\usepackage{graphicx}
\usepackage{multirow}
\usepackage{amsfonts}

\usepackage{amsmath}
\usepackage{pifont}
\usepackage{tabularx}
\usepackage{caption, booktabs}
\usepackage{algorithm}
\usepackage{algpseudocode}
\usepackage{amssymb}
\usepackage{setspace}
\usepackage{bbm}
\usepackage{arydshln}

\makeatletter
\newcommand\footnoteref[1]{\protected@xdef\@thefnmark{\ref{#1}}\@footnotemark}
\makeatother
\newcolumntype{P}[1]{>{\centering\arraybackslash}p{#1}}

\usepackage{xspace}
\newcommand{\oursfull}[0]{Plan-as-query Example Retrieval for few-shot prompting in Code generation\xspace}
\newcommand{\ours}[0]{\textsc{PERC}\xspace}
\usepackage{graphicx}
\usepackage{adjustbox}
\usepackage{subfig}
\makeatletter
\newcommand{\thickhline}{
    \noalign {\ifnum 0=`}\fi \hrule height 1pt
    \futurelet \reserved@a \@xhline
}
\newcolumntype{"}{@{\hskip\tabcolsep\vrule width 1pt\hskip\tabcolsep}}
\makeatother
\usepackage{mathabx}
\usepackage{amsfonts}

\makeatletter
\newcommand*{\blackleq}{
  \mathrel{
    \mathpalette\@blackleq{}
  }
}
\newcommand*{\@blackleq}[2]{
  \vcenter{
    \m@th
    \setbox0=\hbox{$#1\mkern3mu$}
    \setbox2=\hbox{$#1\vcenter{}$}
    \setbox4=\hbox{\raisebox{-\ht2}[.2pt][.2pt]{$#1-$}}
    \hbox{$#1\blacktriangleleft$}
    \nointerlineskip
    \kern\wd0 
    \copy4 
  }
}
\makeatother
\usepackage{dsfont}

\usepackage{multirow}
\usepackage{kotex} 

\usepackage{mathtools}

\definecolor{my_blue}{RGB}{0,112,192}

\usepackage{placeins}

\usepackage{tikz}

\title{\ours: Plan-As-Query Example Retrieval for \\Underrepresented Code Generation}

\author{
Jaeseok Yoo\textsuperscript{1}\quad
Hojae Han\textsuperscript{1}\quad
Youngwon Lee\textsuperscript{1}\quad
Jaejin Kim\textsuperscript{1}\textsuperscript{2}\quad
Seung-won Hwang\textsuperscript{1}\textsuperscript{2}\thanks{Corresponding author.}\\
\mbox{}\textsuperscript{1}Seoul National University\\
\mbox{}\textsuperscript{2}Interdisciplinary Program in Artificial Intelligence, Seoul National University\\
\texttt{\string{jaeseok2.yoo,stovecat,ludaya,jaejin.kim,seungwonh\string}@snu.ac.kr}
}

\begin{document}
\maketitle

\begin{abstract}
Code generation with large language models has shown significant promise, especially when employing retrieval-augmented generation (RAG) with few-shot examples. However, selecting effective examples that enhance generation quality remains a challenging task, particularly when the target programming language (PL) is underrepresented. 
In this study, we present two key findings: (1) retrieving examples whose presented algorithmic plans can be referenced for generating the desired behavior significantly improves generation accuracy, and (2) converting code into pseudocode effectively captures such algorithmic plans, enhancing retrieval quality even when the source and the target PLs are different. Based on these findings, we propose \oursfull (\ours), a novel framework that utilizes algorithmic plans to identify and retrieve effective examples. We validate the effectiveness of \ours through extensive experiments on the CodeContests, HumanEval and MultiPL-E benchmarks:
\ours consistently outperforms the state-of-the-art RAG methods in code generation, both when the source and target programming languages match or differ, highlighting its adaptability and robustness in diverse coding environments.
\end{abstract}

\begin{figure}[t]
\centering
\hspace*{-0.3cm}
\begin{tabular}{cc}
\includegraphics[width=1.0\linewidth]{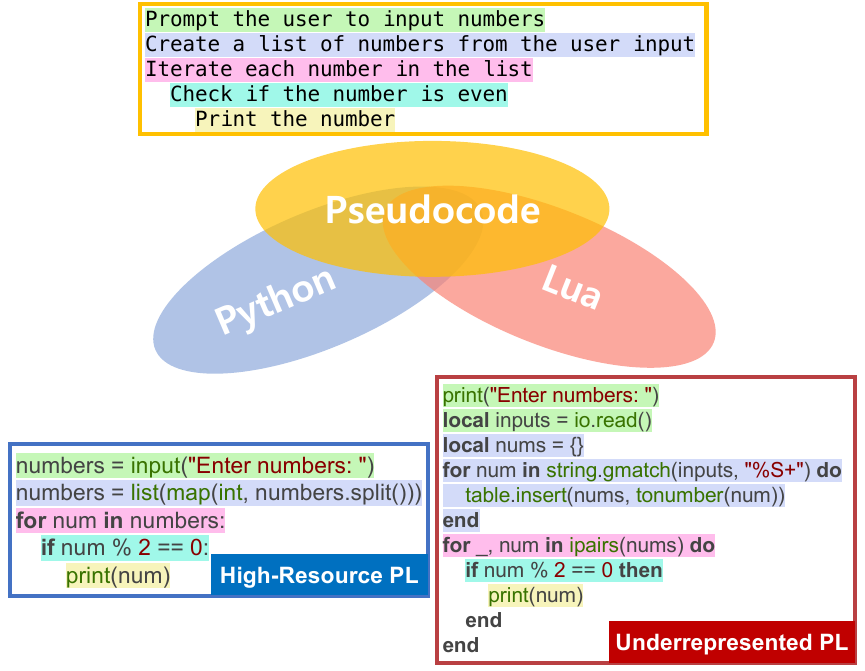}
\end{tabular}
\caption{Two Python and Lua code snippets have different modalities but implement the same algorithmic plans. \ours uses pseudocode describing the algorithmic plans to minimize noise from modality differences. Colors in code represent equivalent steps.}
\label{fig:intro_fig}
\end{figure}

\section{Introduction}
\label{intro}

Code generation using large language models (LLMs) has shown significant potential, particularly when retrieval-augmented generation (RAG)  with few-shot examples is employed~\cite{parvez-etal-2021-retrieval-augmented, 10172590, zhang-etal-2023-repocoder}. 
However, selecting effective examples to improve code generation quality remains a challenging task. 
This is even more difficult when the target programming language (PL) is underrepresented, as the construction of the few-shot example pool for retrieval is non-trivial.

To construct the retrieval pool for an underrepresented PL, we can transfer the retrieval pool from a high-resource PL.
However, this in turn interferes with the state-of-the-art few-shot prompting approaches in code generation~\cite{10172590,zhang-etal-2023-repocoder}, which employ code to retrieve examples. 
Figure~\ref{fig:intro_fig} illustrates such an example; although the Python code on the lower left and the Lua code on the lower right follow the same algorithmic steps, both lexical-based~\cite{Robertson2009ThePR} and embedding-based~\citep{song2020mpnet} retrieval fall short in capturing their algorithmic similarity due to syntactic and structural difference~\citep{an-etal-2023-skill}.

To overcome this, we propose \oursfull (\ours). \ours leverages algorithmic plans such as pseudocode to retrieve examples.
Converting code to pseudocode reduces syntactic noise and captures algorithmic similarity, enhancing retrieval quality across programming languages.
Also, such plans can aid generation by participating in reasoning chains, further improving the generation accuracy.


We demonstrate \ours's effectiveness in two key scenarios: First, plan-based example selection improves code generation accuracy for same-language tasks on CodeContests~\cite{li2022competition} and HumanEval~\cite{codex}. Second, on MultiPL-E~\cite{cassano2023multipl}, \ours enhances code generation for underrepresented languages by leveraging data from high-resource languages.

Our contribution is three-fold:
\begin{itemize}
\item We propose \ours, a novel framework of leveraging algorithmic plans for few-shot example retrieval in code generation.
\item We demonstrate that plan-based retrieval improves same-language code generation on competitive programming and general-purpose coding tasks.
\item We confirm that \ours can leverage high-resource PLs to improve code generation accuracy in underrepresented PLs.
\end{itemize}

\section{Related Work}
\label{related}

\paragraph{Retrieval-Augmented Code Generation} 

Previous works adopting RAG in code generation tasks have primarily focused on enhancing the accuracy of generated code by retrieving from the target PL pool~\citep{parvez-etal-2021-retrieval-augmented, lu-etal-2022-reacc}.
More recently, CEDAR~\cite{10172590} retrieved few-shot examples based on code-code similarity, and RepoCoder~\cite{zhang-etal-2023-repocoder} leveraged LLM-generated code snippets in target PL to expand queries, allowing for improved retrieval.

\paragraph{Utilizing Algorithmic Plans in Code Retrieval and Generation} \citet{pmlr-v157-han21a} viewed pseudocode as algorithmic plan of code, to reduce the modality gap between text and code in code search task. \citet{10.1145/3672456} used pseudocode-based algorithmic plans for code generation through few-shot prompting, and \citet{sun-etal-2024-unicoder} used pseudocode to bridge different programming languages.

\paragraph{Our distinction.}
We are the first to leverage algorithmic plans in retrieval-augmented code generation, which allows to retrieve effective few-shot examples by reducing lexical bias. As a result, our proposed \ours naturally adapts to source-target PLs mismatches.
\begin{figure}[t]
\centering  
\includegraphics[width=1.0\linewidth]{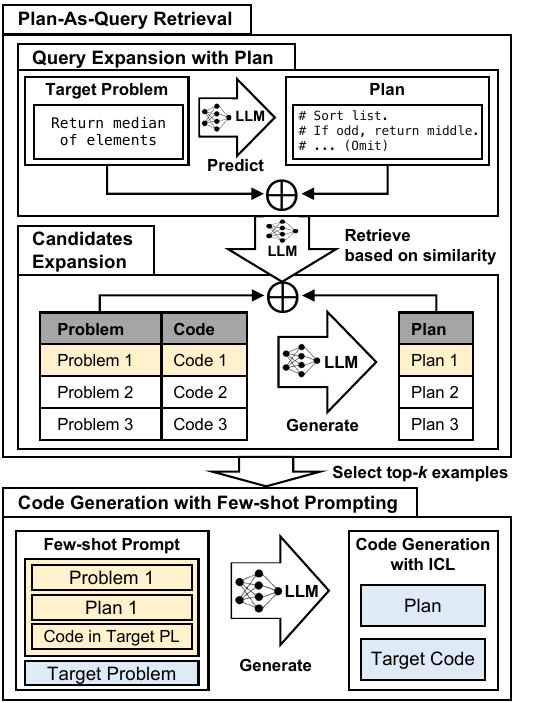}
\caption{Overview of \ours: (1) plan-based retrieval and (2) code generation with few-shot prompting.
\ours retrieves examples with the most similar algorithmic plans. Yellow and blue respectively signifies the triplet of the selected few-shot example and the target problem.}
\label{fig:overview}
\end{figure}

\section{Preliminaries}

Given a natural language query $t_q$ describing a desired program, code generation aims to return the corresponding implementation. 
In few-shot example retrieval, we draw relevant text-code pairs $(t,c)$ from an example pool $\mathcal{P}$ to supplement LLM's knowledge and guide generation.




\paragraph{Problem-As-Query}
A baseline approach maps query $t_q$ and text descriptions $t$ into a shared latent space using encoder $\psi$. The top-$k$ examples are selected based on similarity $\mathrm{sim}(\cdot)$:
\begin{equation}
E = \mathrm{topk}_{{(t,c)}\in \mathcal{P}}\ \mathrm{sim} (\psi(t_q), \psi(t) ) ,
\end{equation}
where $E$ is the set of indices of chosen examples.

\paragraph{CEDAR} 
\citet{10172590} selects examples based on code-code similarity. As the prompt lacks code for querying, we use the LLM's predicted code $\hat{c}_q$ from $t_q$ to retrieve examples:
\begin{equation}
E_{\textrm{CD}} = \mathrm{topk}_{(t,c)\in \mathcal{P}} \ \mathrm{sim} (\psi(\hat{c}_q), \psi(c) ) .
\end{equation}

\paragraph{RepoCoder} \citet{zhang-etal-2023-repocoder} generates code $\hat{c}_q$ from $t_q$ and expands the query, following LLM-based query expansion~\cite{wang-etal-2023-query2doc}. The retrieval combines both problem description and code:
\begin{equation} \label{eq:repocoder}
E_{\textrm{RC}} = \mathrm{topk}_{(t,c)\in \mathcal{P}} \ \mathrm{sim} (\psi(t_q; \hat{c}_q), \psi(t; c) ) ,
\end{equation} 
where semicolon denotes concatenation.


\newcolumntype{L}[1]{>{\raggedright\arraybackslash}p{#1}}
\newcolumntype{g}{>{\columncolor{gray!10}}c}
\begin{table*}[t]
    \centering
    \resizebox{0.98\textwidth}{!}{%
    \begin{tabular}{lcccccccc} 
        \thickhline
        Benchmark & CodeContests & HumanEval & \multicolumn{6}{c}{MultiPL-E} \\
        Target PL & Python & Python & Rust & Julia & Lua & Ruby & Go & R \\
        Data Availability & High-resource & High-resource &  \multicolumn{2}{c}{High-resource} & \multicolumn{2}{c}{$\xleftrightarrow{\hspace{1cm} \text{} \hspace{1cm}}$} & \multicolumn{2}{c}{Underrepresented}  \\
        \hline
        Few-shot Prompting Method & \multicolumn{8}{c}{Pass@1 (\%)} \\
        \hline
        w/o Examples & 2.72 & 73.17 & 62.82 & 51.45 & 49.94 & 47.76 & 30.91 & 14.78 \\
        Random Selection & 5.82 & 72.56 & 61.86 & 52.77 & 60.68 & \underline{68.01} & 45.58 & 33.42 \\
        Problem-As-Query Retrieval & 4.97 & 69.51 & \underline{63.14} & 53.65 & 60.93 & 65.22 & \underline{75.06} & 31.06 \\
        CEDAR & 5.82 & 72.44 & 62.63 & \underline{53.71} & \underline{61.61} & 61.68 & \underline{75.06} & 32.67 \\
        RepoCoder & \underline{6.48} & \underline{73.78} & 62.37 & 52.83 & 60.81 & 67.27 & 71.75 & \underline{34.16} \\
        \rowcolor{gray!10}
        \ours & \textbf{6.61} & \textbf{76.04} & \textbf{63.78} & \textbf{54.21} & \textbf{64.10} & \textbf{69.81} & \textbf{76.49} & \textbf{34.35} \\
        \thickhline
    \end{tabular}
    }
    \normalsize
    \caption{Pass@1 scores of GPT-3.5-Turbo-16k augmented with different strategies for retrieving few-shot examples from Python source pools across CodeContests, HumanEval, and MultiPL-E benchmarks.
    Boldface indicates the best values while underline indicates the second-highest accuracy.}
    \label{tab:main}
\end{table*}
\newcolumntype{L}[1]{>{\raggedright\arraybackslash}p{#1}}
\newcolumntype{g}{>{\columncolor{gray!10}}c}
\begin{table*}[t]
    \centering
    \resizebox{0.98\textwidth}{!}{%
    \begin{tabular}{lcccccccc} 
        \thickhline
        Benchmark & CodeContests & HumanEval & \multicolumn{6}{c}{MultiPL-E} \\
        Target PL & Python & Python & Rust & Ruby & Lua & Julia & Go & R \\
        Data Availability & High-resource & High-resource &  \multicolumn{2}{c}{High-resource} & \multicolumn{2}{c}{$\xleftrightarrow{\hspace{1cm} \text{} \hspace{1cm}}$} & \multicolumn{2}{c}{Underrepresented}  \\
        \hline
        Few-shot Prompting Method & \multicolumn{8}{c}{Pass@1 (\%)} \\
        \hline
        w/o Examples & 4.97 & 87.87 & 81.22 & 78.88 & 66.96 & 61.51 & 52.73 & 49.07 \\
        Random Selection & 5.58 & 86.59 & 81.35 & 79.50 & 73.98 & 68.36 & 47.01 & 53.42 \\
        Problem-As-Query Retrieval & 7.64 & 87.07 & 80.58 & 80.12 & 72.92 & 65.97 & 68.83 & 52.80 \\
        CEDAR & \underline{8.18} & \textbf{88.17} & 81.15 & 80.75 & 74.04 & \underline{69.37} & \underline{71.43} & 53.42 \\
        RepoCoder & 7.33 & 86.46 & \underline{81.54} & \underline{81.99} & \textbf{75.65} & 67.42 & 70.71 & \underline{55.28} \\
        \rowcolor{gray!10}
        \ours & \textbf{8.48} & \textbf{88.17} & \textbf{82.95} & \textbf{83.85} & \underline{75.22} & \textbf{70.69} & \textbf{71.69} & \textbf{57.14} \\ 
        \thickhline
    \end{tabular}
    }
    \normalsize
    \caption{Pass@1 scores of GPT-4o-mini augmented with different strategies for retrieving few-shot examples from Python source pools across CodeContests, HumanEval, and MultiPL-E benchmarks.}
    \label{tab:gpt-4o-mini}
\end{table*}

\section{\ours}
\ours retrieves relevant examples using algorithmic plans in pseudocode. These plans capture high-level logic while minimizing cross-lingual lexical differences, thereby supporting accurate code generation.

As depicted in Figure~\ref{fig:overview}, the workflow of \ours consists of two key steps. First, it drafts a plan for the given problem to form an expanded query, which is used to retrieve examples that were projected to plan space in indexing time. Then, the retrieved examples and their plans are integrated into a reasoning chain to generate a revised plan and the final code.

\subsection{Plan-As-Query Example Retrieval}

A key contribution of
\ours is the use of algorithmic plans written in pseudocode, for effective retrieval.
Specifically, an LLM generates pseudocode $\hat{p}$ for the retrieval pool $\mathcal{P}$ offline, and $\hat{p}_q$ for $t_q$ at inference time.
Then, the query is expanded with $\hat{p}_q$ as follows:
\begin{equation} \label{eq:ours}
E_{\textrm{PERC}} = \mathrm{topk}_{(t,\hat{p},c)} \ \mathrm{sim} (\psi(t_q; \hat{p}_q), \psi(t; \hat{p}) ),
\end{equation} 
where the in-context example for $\hat{p}_q$ is provided in Appendix~\ref{icl_examples}.
As illustrated in Figure~\ref{fig:intro_fig}, Eq (\ref{eq:ours}) avoids surface-level details by projecting code $c$ into plans. This is in contrast to Eq (\ref{eq:repocoder}), which exposes the retriever to such distractions, especially when $\hat{c}$ and $c$ use different PLs.

\subsection{Code Generation with Examples}
\label{subsec:prompting}
We use generated pseudocode as intermediate reasoning steps for code generation~\cite{10.1145/3672456}.
Each few-shot example in our prompt consists of a triplet $(t, \hat{p}, c)$, where text description $t$, generated pseudocode $\hat{p}$, and code $c$ guide the LLM to utilize pseudocode in its reasoning chain:
\begin{equation}
\texttt{prompt} = [[ t; \hat{p}; c ]_{(t,\hat{p},c)\in E_{\textrm{PERC}}} ; t_q].
\end{equation}

When the target programming language differs from that of example code $c$, we replace $c$ with generated code $\hat{c}$ in the target language, where the LLM generates $\hat{c}$ using the in-context example shown in Appendix~\ref{icl_examples}:
\begin{equation}
\texttt{prompt} = [[t; \hat{p}; \hat{c}]_{(t,\hat{p},c)\in E_{\textrm{PERC}}} ; t_q]. 
\end{equation}

\section{Experiments}

\subsection{Experimental Setup}
Experiments were conducted using GPT-3.5-Turbo-16k and GPT-4o-mini as the backbone LLMs. Other implementation details regarding the embedding-based retrieval and hyperparameter configuration for code generation can be found in Appendix~\ref{impl_detail}.

\paragraph{Metrics} We evaluated the performance of \ours using the widely used Pass@1 metric~\cite{codex}, which is an unbiased estimator of the model's chance of producing a correct code sample in a single attempt.

\paragraph{Baselines} 
We compared our method against several established baselines to highlight the effectiveness of \ours: 1) \textbf{w/o Examples} generates code directly without using few-shot examples, 2) \textbf{Random Selection} uses a randomly chosen, then fixed set of examples, 3) \textbf{Problem-As-Query Retrieval} retrieves examples based on problem-problem similarity, 4) \textbf{CEDAR}~\cite{10172590} uses code-code similarity, and 5) \textbf{RepoCoder}~\cite{zhang-etal-2023-repocoder} expands the query with predicted code.

\paragraph{Datasets} For evaluation, we used CodeContests~\cite{li2022competition}, HumanEval~\cite{codex}, and MultiPL-E (HumanEval;~\citealp{cassano2023multipl}) benchmarks.
For CodeContests, we used the train split as the example pool, while MBPP~\cite{austin2021programsynthesislargelanguage} benchmark was used for the other two. Throughout the benchmarks, we used Python—a high-resource PL—as the source. 
We used Python as the primary target PL since it is the only officially supported language in both CodeContests and HumanEval benchmarks. For additional target PLs, we selected languages based on the frequency classes (NICHE, LOW, MEDIUM) established in MultiPL-E. We randomly chose two PLs from each class: Ruby and Go (MEDIUM), Rust and R (LOW), and Lua and Julia (NICHE). 

\subsection{RAG from Same PL Pool: CodeContests and HumanEval}
Table~\ref{tab:main} shows that \ours outperforms all baselines, achieving Pass@1 scores of 6.61\% and 76.04\% on CodeContests and HumanEval, respectively, using the GPT-3.5-Turbo-16k model. Similarly, the results for GPT-4o-mini in Table~\ref{tab:gpt-4o-mini} show Pass@1 scores of 8.48\% and 88.17\%, demonstrating consistently high performance across benchmarks. 
This supports that retrieval based on algorithmic plans better captures the logic of the code and allows to surface more effective demonstrations in top-$k$.

\subsection{RAG from Cross-PL Pool: MultiPL-E}
The results for each PL in MultiPL-E, presented in Tables~\ref{tab:main} and~\ref{tab:gpt-4o-mini}, are sorted in descending order of code generation accuracy without examples. PLs with higher accuracy are considered to have higher data availability, while those with lower accuracy are regarded as underrepresented.

Using the GPT-3.5-Turbo-16k model, \ours achieves the best Pass@1 scores across all PLs, with notable results such as 69.81\% for Ruby, 63.78\% for Rust, and 64.10\% for Lua, as shown in Table~\ref{tab:main}. 
The results for GPT-4o-mini, presented in Table~\ref{tab:gpt-4o-mini}, also emphasizes \ours's effectiveness, showing consistent Pass@1 score improvements, including 83.85\% for Ruby, 70.69\% for Julila, and 57.14\% for R.

By effectively transferring knowledge from high-resource PLs, \ours demonstrated improved code generation accuracy for different PLs, showing its ability to bridge knowledge gaps across PLs with different data availability. 
In contrast, state-of-the-art approaches RepoCoder and CEDAR struggled with code generation. This limitation stemmed from their reliance on code-based retrieval, where modality differences introduced noise and hindered the identification of algorithmically relevant code.\footnote{One may consider a cost-exhaustive approach of translating all the code in the pool to target (underrepresented) PLs, which incurs $\mathcal{O}(|\mathcal{T}||\mathcal{P}|)$ cost where $\mathcal{T}$ is the set of target PLs to handle, whereas \ours only requires $\mathcal{O}(|\mathcal{P}|)$.}

\section{Analysis and Discussion}

\paragraph{Open-Source LLM as a Backbone}
Table~\ref{tab:llama-3.1} shows consistent accuracy improvements with the open-source model Llama-3.1-8B-Instruct. \ours outperformed the baselines and demonstrated effective performance across PLs like Ruby, Lua, and R in the MultiPL-E benchmark. This highlights the improvements with \ours generalizes well to smaller, public backbone models.

\newcolumntype{L}[1]{>{\raggedright\arraybackslash}p{#1}}
\newcolumntype{g}{>{\columncolor{gray!10}}c}
\begin{table}[t]
    \small
    \centering
    \resizebox{0.4\textwidth}{!}{%
    \begin{tabular}{lccc} 
        \thickhline
        Benchmark & \multicolumn{3}{c}{MultiPL-E} \\
        Target PL & Ruby & Lua  & R \\
        \hline
        Method & \multicolumn{3}{c}{Pass@1 (\%)} \\
        \hline
        w/o Examples & 46.09 & 39.63  & 18.82 \\
        Random Selection & 45.34 & 38.70 & \underline{22.36} \\ 
        RepoCoder & \underline{46.21} & \underline{41.06} & 18.94 \\
        \rowcolor{gray!10}
        \ours & \textbf{47.33} & \textbf{44.22} & \textbf{23.66} \\ 
        \thickhline
    \end{tabular}
    }
    \normalsize
    \caption{Pass@1 scores of Llama-3.1-8B-Instruct augmented with different strategies for retrieving few-shot examples on Python source pool, on MultiPL-E benchmarks.}
    \label{tab:llama-3.1}
\end{table}
\newcolumntype{g}{>{\columncolor{gray!10}}c}
\begin{table}[t]
    \small
    \centering
    \begin{tabular}{lcccc}
        \thickhline
         Benchmark & \multicolumn{2}{c}{CodeContests} & \multicolumn{2}{c}{MultiPL-E} \\
        Cand. PL & C++ & Java & C++ & Java \\
        Target PL & Python & Python & Lua & Lua \\
        \hline
        RepoCoder & \phantom{0}5.45 & \phantom{0}5.94 & 58.88 &	58.01 \\
        \rowcolor{gray!10}
        \ours & \phantom{0}\textbf{6.61} & \phantom{0}\textbf{6.06} & \textbf{64.60} & \textbf{64.60}  \\ 
        \thickhline
    \end{tabular}%
    \normalsize
    \caption{Pass@1 scores of \ours and RepoCoder when using C++ and Java candidates in the CodeContests and MultiPL-E Lua benchmarks.}
    \label{tab:codecontests_cpp_java}
\end{table}
\begin{table}[t!]
    \centering
    \resizebox{.48\textwidth}{!}{%
    \begin{tabular}{lccc}
        \thickhline
         Cand. PL & Python & Python/C++ & Python/C++/Java  \\
         \hline
        RepoCoder & 6.48 & 5.88 & 4.85 \\
        \rowcolor{gray!10}
        \ours & \textbf{6.61} & \textbf{6.48} & \textbf{6.12} \\
        \thickhline
    \end{tabular}%
    }
    \normalsize
    \caption{Pass@1 scores on CodeContests as more examples from different PLs are added to the retrieval pool.}
    \label{tab:cand_pls}
\end{table}
\begin{figure}[t!]
\centering
\hspace*{-0.3cm}
\begin{tabular}{cc}
\includegraphics[width=1.0\linewidth]{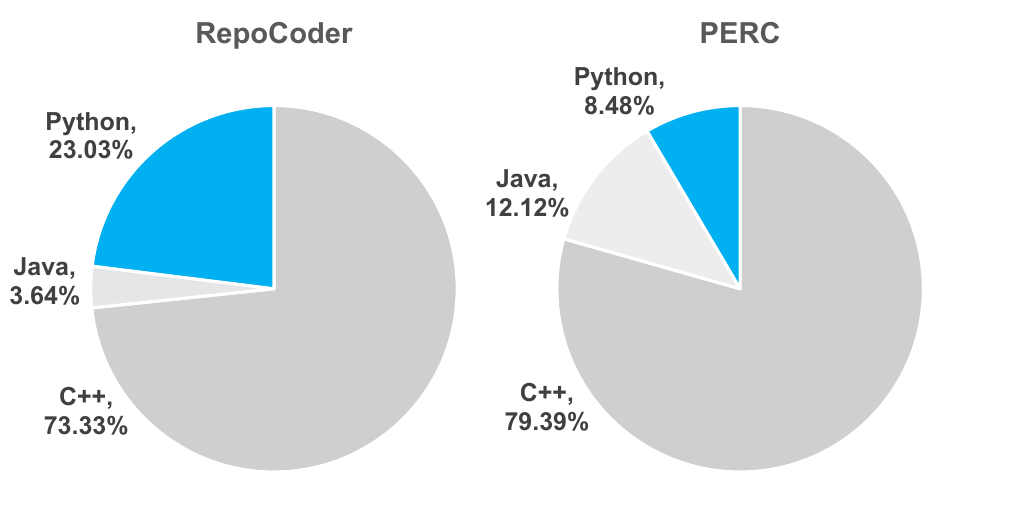}
\end{tabular}
\caption{Source PL distribution of retrieved examples on CodeContests, under the Mixed PL Pool setting of Table~\ref{tab:cand_pls}.
The target PL, Python, is highlighted in blue.
}
\label{fig:pl_dist}
\end{figure}
\begin{table}[t]
    \small
    \centering
    \begin{tabular}{lc}
        \thickhline
        Method & Pass@$1$ (\%)  \\
        \hline
        \rowcolor{gray!10}
        \ours & \textbf{64.10} \\
        -\phantom{ }Converting to Target PL & 55.28 \\
        \thickhline
    \end{tabular}
    \normalsize
    \caption{Pass@1 of \ours on MultiPL-E-Lua benchmark with and without code conversion to target PL.}
    \label{tab:convert_pls}
\end{table}
\begin{table}[t!]
    \small
    \centering
    \begin{tabular}{lc}
        \thickhline
        Method & Pass@$1$ (\%)  \\
        \hline
        \ours w/ Converted Code & 5.52 \\
        \ours w/ Gold Code & 5.70 \\
        \thickhline
    \end{tabular}
    \normalsize
    \caption{Pass@1 difference when replacing generated target PL code with gold target PL code on subsets of example pools containing both C++ and Python code in CodeContets}
    \label{tab:correct_convert_pls}
\end{table}

\paragraph{Using C++ and Java as Source PLs}
Table~\ref{tab:codecontests_cpp_java} shows that \ours is also effective when using source PLs other than Python,
namely C++ and Java.
These results demonstrate that selecting examples based on algorithmic plans, regardless of the source PLs, can enhance code generation accuracy. The accuracy improvements using C++ and Java code pool for all MultiPL-E benchmarks are detailed in Appendix~\ref{cpp_java_multipl_e}.

\paragraph{Mixed PL Pool}
\label{subsec:mixing}

As shown in Table~\ref{tab:cand_pls}, when C++ and Java code were incrementally added to the retrieval pool, \ours maintained higher accuracy while RepoCoder suffered from severe performance degradation.
Again, this showcases the adaptability and robustness of retrieving examples with plans, than with code. 

As illustrated in Figure~\ref{fig:pl_dist}, \ours outperforms RepoCoder while retrieving less examples of the target PL, Python; this empirically supports that \ours retrieves useful examples in non-target PLs.

\paragraph{Converting Code to Target PL}
As illustrated in Section~\ref{subsec:prompting}, \ours uses the converted code $\hat{c}$ rather than the original code of the retrieved example $c$, if the source and target PLs do not match.
Table~\ref{tab:convert_pls} shows such conversion is crucial,
as syntax, APIs, and other elements specific to the source PL may inadvertently influence the generation and lead to performance degradation. Additionally, as shown in Table~\ref{tab:correct_convert_pls}, replacing the generated target PL code with the gold target PL code in CodeContests,\footnote{The subset of examples with both C++ and Python GT code available was used for evaluation.} resulted in a minimal Pass@1 difference. This indicates potential errors that can be introduced in conversion to target PL are negligible.

\section{Conclusion}
We presented \ours,
a novel framework for code generation that utilizes algorithmic plans both at indexing and generation time to select more effective few-shot examples to guide LLM.
\ours demonstrates notable improvements in Pass@1 on CodeContests and HumanEval benchmark which represent scenario where the target PL is of high-resource, and also on MultiPL-E benchmarks for targeting underrepresented PLs.

\section*{Limitations}
While \ours brings notable performance improvements in code generation with few-shot prompting, even if the source and target PLs do not match, our observation in Section~\ref{subsec:mixing} is that \ours shows slight performance drop as more and more programming languages are added to the retrieval pool.
With an ideal retrieval system suited to selecting the most effective examples, one should observe monotonically increasing performance as more candidates are added to the pool; we leave more investigation and improvements as future work.

\section*{Acknowledgments}
The first author was supported by Samsung Electronics. This work was partly supported by Electronics and Telecommunications Research Institute (ETRI) grant funded by ICT R\&D program of MSIT/IITP (2022-0-00995, Automated reliable source code generation from natural language descriptions), and
the MSIT (Ministry of Science and ICT), Korea, under the ITRC (Information Technology Research Center) support program (IITP-2024-2020-0-01789) supervised by the IITP (Institute for Information \& Communications Technology Planning \& Evaluation). 


\bibliography{anthology_part2,custom}

\begin{thebibliography}{15}
\providecommand{\natexlab}[1]{#1}

\bibitem[{An et~al.(2023)An, Zhou, Lin, Fu, Chen, Zheng, Chen, and Lou}]{an-etal-2023-skill}
Shengnan An, Bo~Zhou, Zeqi Lin, Qiang Fu, Bei Chen, Nanning Zheng, Weizhu Chen, and Jian-Guang Lou. 2023.
\newblock \href {https://doi.org/10.18653/v1/2023.emnlp-main.831} {Skill-based few-shot selection for in-context learning}.
\newblock In \emph{Proceedings of the 2023 Conference on Empirical Methods in Natural Language Processing}, pages 13472--13492, Singapore. Association for Computational Linguistics.

\bibitem[{Austin et~al.(2021)Austin, Odena, Nye, Bosma, Michalewski, Dohan, Jiang, Cai, Terry, Le, and Sutton}]{austin2021programsynthesislargelanguage}
Jacob Austin, Augustus Odena, Maxwell Nye, Maarten Bosma, Henryk Michalewski, David Dohan, Ellen Jiang, Carrie Cai, Michael Terry, Quoc Le, and Charles Sutton. 2021.
\newblock \href {https://arxiv.org/abs/2108.07732} {Program synthesis with large language models}.
\newblock \emph{Preprint}, arXiv:2108.07732.

\bibitem[{Cassano et~al.(2023)Cassano, Gouwar, Nguyen, Nguyen, Phipps-Costin, Pinckney, Yee, Zi, Anderson, Feldman, Guha, Greenberg, and Jangda}]{cassano2023multipl}
Federico Cassano, John Gouwar, Daniel Nguyen, Sydney Nguyen, Luna Phipps-Costin, Donald Pinckney, Ming-Ho Yee, Yangtian Zi, Carolyn~Jane Anderson, Molly~Q Feldman, Arjun Guha, Michael Greenberg, and Abhinav Jangda. 2023.
\newblock \href {https://doi.org/10.1109/TSE.2023.3267446} {Multipl-e: A scalable and polyglot approach to benchmarking neural code generation}.
\newblock \emph{IEEE Transactions on Software Engineering}, 49(7):3675--3691.

\bibitem[{Chen et~al.(2021)Chen, Tworek, Jun, Yuan, de~Oliveira~Pinto, Kaplan, Edwards, Burda, Joseph, Brockman, Ray, Puri, Krueger, Petrov, Khlaaf, Sastry, Mishkin, Chan, Gray, Ryder, Pavlov, Power, Kaiser, Bavarian, Winter, Tillet, Such, Cummings, Plappert, Chantzis, Barnes, Herbert{-}Voss, Guss, Nichol, Paino, Tezak, Tang, Babuschkin, Balaji, Jain, Saunders, Hesse, Carr, Leike, Achiam, Misra, Morikawa, Radford, Knight, Brundage, Murati, Mayer, Welinder, McGrew, Amodei, McCandlish, Sutskever, and Zaremba}]{codex}
Mark Chen, Jerry Tworek, Heewoo Jun, Qiming Yuan, Henrique~Pond{\'{e}} de~Oliveira~Pinto, Jared Kaplan, Harrison Edwards, Yuri Burda, Nicholas Joseph, Greg Brockman, Alex Ray, Raul Puri, Gretchen Krueger, Michael Petrov, Heidy Khlaaf, Girish Sastry, Pamela Mishkin, Brooke Chan, Scott Gray, Nick Ryder, Mikhail Pavlov, Alethea Power, Lukasz Kaiser, Mohammad Bavarian, Clemens Winter, Philippe Tillet, Felipe~Petroski Such, Dave Cummings, Matthias Plappert, Fotios Chantzis, Elizabeth Barnes, Ariel Herbert{-}Voss, William~Hebgen Guss, Alex Nichol, Alex Paino, Nikolas Tezak, Jie Tang, Igor Babuschkin, Suchir Balaji, Shantanu Jain, William Saunders, Christopher Hesse, Andrew~N. Carr, Jan Leike, Joshua Achiam, Vedant Misra, Evan Morikawa, Alec Radford, Matthew Knight, Miles Brundage, Mira Murati, Katie Mayer, Peter Welinder, Bob McGrew, Dario Amodei, Sam McCandlish, Ilya Sutskever, and Wojciech Zaremba. 2021.
\newblock \href {https://arxiv.org/abs/2107.03374} {Evaluating large language models trained on code}.
\newblock \emph{CoRR}, abs/2107.03374.

\bibitem[{Han et~al.(2021)Han, Lee, Kim, and Hwang}]{pmlr-v157-han21a}
Hojae Han, Youngwon Lee, Minsoo Kim, and Seung-won Hwang. 2021.
\newblock \href {https://proceedings.mlr.press/v157/han21a.html} {Bridging code-text representation gap using explanation}.
\newblock In \emph{Proceedings of The 13th Asian Conference on Machine Learning}, volume 157 of \emph{Proceedings of Machine Learning Research}, pages 1033--1048. PMLR.

\bibitem[{Jiang et~al.(2024)Jiang, Dong, Wang, Zheng, Shang, Li, Jin, and Jiao}]{10.1145/3672456}
Xue Jiang, Yihong Dong, Lecheng Wang, Fang Zheng, Qiwei Shang, Ge~Li, Zhi Jin, and Wenpin Jiao. 2024.
\newblock \href {https://doi.org/10.1145/3672456} {Self-planning code generation with large language models}.
\newblock \emph{ACM Trans. Softw. Eng. Methodol.}

\bibitem[{Li et~al.(2022)Li, Choi, Chung, Kushman, Schrittwieser, Leblond, Eccles, Keeling, Gimeno, Lago, Hubert, Choy, de~Masson~d'Autume, Babuschkin, Chen, Huang, Welbl, Gowal, Cherepanov, Molloy, Mankowitz, Robson, Kohli, de~Freitas, Kavukcuoglu, and Vinyals}]{li2022competition}
Yujia Li, David Choi, Junyoung Chung, Nate Kushman, Julian Schrittwieser, R{\'e}mi Leblond, Tom Eccles, James Keeling, Felix Gimeno, Agustin~Dal Lago, Thomas Hubert, Peter Choy, Cyprien de~Masson~d'Autume, Igor Babuschkin, Xinyun Chen, Po-Sen Huang, Johannes Welbl, Sven Gowal, Alexey Cherepanov, James Molloy, Daniel~J. Mankowitz, Esme~Sutherland Robson, Pushmeet Kohli, Nando de~Freitas, Koray Kavukcuoglu, and Oriol Vinyals. 2022.
\newblock \href {https://doi.org/10.1126/science.abq1158} {Competition-level code generation with alphacode}.
\newblock \emph{Science}, 378(6624):1092--1097.

\bibitem[{Lu et~al.(2022)Lu, Duan, Han, Guo, Hwang, and Svyatkovskiy}]{lu-etal-2022-reacc}
Shuai Lu, Nan Duan, Hojae Han, Daya Guo, Seung-won Hwang, and Alexey Svyatkovskiy. 2022.
\newblock \href {https://doi.org/10.18653/v1/2022.acl-long.431} {{R}e{ACC}: A retrieval-augmented code completion framework}.
\newblock In \emph{Proceedings of the 60th Annual Meeting of the Association for Computational Linguistics (Volume 1: Long Papers)}, pages 6227--6240, Dublin, Ireland. Association for Computational Linguistics.

\bibitem[{Nashid et~al.(2023)Nashid, Sintaha, and Mesbah}]{10172590}
Noor Nashid, Mifta Sintaha, and Ali Mesbah. 2023.
\newblock \href {https://doi.org/10.1109/ICSE48619.2023.00205} {Retrieval-based prompt selection for code-related few-shot learning}.
\newblock In \emph{2023 IEEE/ACM 45th International Conference on Software Engineering (ICSE)}, pages 2450--2462.

\bibitem[{Parvez et~al.(2021)Parvez, Ahmad, Chakraborty, Ray, and Chang}]{parvez-etal-2021-retrieval-augmented}
Md~Rizwan Parvez, Wasi Ahmad, Saikat Chakraborty, Baishakhi Ray, and Kai-Wei Chang. 2021.
\newblock \href {https://doi.org/10.18653/v1/2021.findings-emnlp.232} {Retrieval augmented code generation and summarization}.
\newblock In \emph{Findings of the Association for Computational Linguistics: EMNLP 2021}, pages 2719--2734, Punta Cana, Dominican Republic. Association for Computational Linguistics.

\bibitem[{Robertson and Zaragoza(2009)}]{Robertson2009ThePR}
Stephen~E. Robertson and Hugo Zaragoza. 2009.
\newblock \href {https://api.semanticscholar.org/CorpusID:207178704} {The probabilistic relevance framework: Bm25 and beyond}.
\newblock \emph{Found. Trends Inf. Retr.}, 3:333--389.

\bibitem[{Song et~al.(2020)Song, Tan, Qin, Lu, and Liu}]{song2020mpnet}
Kaitao Song, Xu~Tan, Tao Qin, Jianfeng Lu, and Tie-Yan Liu. 2020.
\newblock Mpnet: Masked and permuted pre-training for language understanding.
\newblock \emph{Advances in neural information processing systems}, 33:16857--16867.

\bibitem[{Sun et~al.(2024)Sun, Chai, Yang, Yin, Guo, Liu, Wang, Yang, and Li}]{sun-etal-2024-unicoder}
Tao Sun, Linzheng Chai, Jian Yang, Yuwei Yin, Hongcheng Guo, Jiaheng Liu, Bing Wang, Liqun Yang, and Zhoujun Li. 2024.
\newblock \href {https://aclanthology.org/2024.acl-long.100} {{U}ni{C}oder: Scaling code large language model via universal code}.
\newblock In \emph{Proceedings of the 62nd Annual Meeting of the Association for Computational Linguistics (Volume 1: Long Papers)}, pages 1812--1824, Bangkok, Thailand. Association for Computational Linguistics.

\bibitem[{Wang et~al.(2023)Wang, Yang, and Wei}]{wang-etal-2023-query2doc}
Liang Wang, Nan Yang, and Furu Wei. 2023.
\newblock \href {https://doi.org/10.18653/v1/2023.emnlp-main.585} {Query2doc: Query expansion with large language models}.
\newblock In \emph{Proceedings of the 2023 Conference on Empirical Methods in Natural Language Processing}, pages 9414--9423, Singapore. Association for Computational Linguistics.

\bibitem[{Zhang et~al.(2023)Zhang, Chen, Zhang, Keung, Liu, Zan, Mao, Lou, and Chen}]{zhang-etal-2023-repocoder}
Fengji Zhang, Bei Chen, Yue Zhang, Jacky Keung, Jin Liu, Daoguang Zan, Yi~Mao, Jian-Guang Lou, and Weizhu Chen. 2023.
\newblock \href {https://doi.org/10.18653/v1/2023.emnlp-main.151} {{R}epo{C}oder: Repository-level code completion through iterative retrieval and generation}.
\newblock In \emph{Proceedings of the 2023 Conference on Empirical Methods in Natural Language Processing}, pages 2471--2484, Singapore. Association for Computational Linguistics.

\end{thebibliography}

\clearpage
\appendix

\onecolumn

\section{Implementation Details}
\label{impl_detail}
Throughout the experiments, we used GPT-3.5-Turbo-16k and GPT-4o-mini-2024-07-18 as the base LLM; for decoding, we applied nucleus sampling with $p=0.95$ and sharpening with temperature $T = 0.8$.
In all retrieval experiments, we employed semantic search using embedding models for retrieval, namely MPNet~\cite{song2020mpnet}\footnote{\url{https://huggingface.co/sentence-transformers/all-mpnet-base-v2}} as the encoder $\psi$.
For a fair comparison, we also used this retriever for the RepoCoder baseline as well, which originally used a sparse bag-of-words model.
The top-$k$ candidates were then selected based on MPNet-based embeddings using cosine similarity, ensuring consistency across all methods.

In all few-shot prompting-based code generation experiments, ICL was performed using an inference chain consisting of problem description, pseudocode, and code. To ensure fair comparison of retrieval methods, the code was replaced with code converted to the target PL, as done in PERC’s approach.
Regarding the number of shots, we employed 3-shot prompting for all benchmarks, except CodeContests, where 1-shot prompting was necessary due to the model's 16k token limit.

Our code generation implementation primarily relied on the LangChain library.\footnote{\url{https://github.com/langchain-ai/langchain}}
To execute and evaluate the generated code, we used code from the CodeEval repository\footnote{\url{https://huggingface.co/spaces/evaluate-metric/code_eval}} hosted on Huggingface for running evaluating on CodeContests and HumanEval benchmarks. Additionally, we utilized code from the bigcode-evaluation-harness repository\footnote{\url{https://github.com/bigcode-project/bigcode-evaluation-harness}} to evaluate on the MultiPL-E benchmark suite.

\section{Using C++ and Java Candidates for MultiPL-E Benchmarks}

\label{cpp_java_multipl_e}

\newcolumntype{g}{>{\columncolor{gray!10}}c}
\begin{table*}[h]
    \centering
    \begin{tabular}{lcccccc}
        \thickhline
         Benchmark & \multicolumn{6}{c}{MultiPL-E} \\
        Candidate PL & C++ & Java & C++ & Java & C++ & Java \\
        Target PL & Lua & Lua & Ruby & Ruby & R & R\\
        \hline
        RepoCoder & 58.88 &	58.01 & 68.51 &	65.28 & 30.43 &	33.91 \\
        \rowcolor{gray!10}
        \ours & \textbf{64.60} & \textbf{64.60} & \textbf{71.18} & \textbf{66.15} & \textbf{31.68}	& \textbf{34.66} \\ 
        \thickhline
    \end{tabular}
    \normalsize
    \caption{Experimental results comparing the Pass@1 of PERC and RepoCoder when using C++ and Java candidates in the MultiPL-E HumanEval-Lua, Ruby, and R benchmarks.}
    \label{tab:multiPL_cpp_java}
\end{table*}

As shown in Table~\ref{tab:multiPL_cpp_java}, even when the candidate programming languages are C++ and Java rather than Python, \ours outperforms RepoCoder in the MultiPL-E HumanEval-Lua, Ruby, and R benchmarks. These results demonstrate that selecting examples based on pseudocode, regardless of the candidate PLs, can improve code generation accuracy for underrepresented PLs by leveraging the knowledge from examples written in high-resource PLs.

\clearpage
\section{Ablation Studies}
\subsection{Pseudocode as Reasoning Chain}
\label{appendix:pseudocode}
\begin{table}[h]
    \centering
    \begin{tabular}{lccc} 
        \thickhline
        Benchmark & \multicolumn{2}{c}{MultiPL-E Lua} \\
        Reasoning Chain &  - & Pseudocode \\
        \hline
        w/o Examples & 49.94 & - \\
        Random Selection & 59.38 & \textbf{60.68} \\ 
        Problem-As-Query Retrieval & 60.81 & \textbf{60.93} \\
        CEDAR & 59.69 & \textbf{61.61} \\
        RepoCoder & 58.94 & \textbf{60.81} \\
        \rowcolor{gray!10}
        \ours & 61.99 & \textbf{64.10} \\ 
        \thickhline
    \end{tabular}
    \normalsize
    \caption{Pass@1 scores of GPT-3.5-Turbo-16k with and without pseudocode as a reasoning chain.}
    \label{tab:pseudo}
\end{table}
\subsection{Different Retrieval Methods}
\label{appendix:retriever}
\begin{table}[h]
    \centering
    \begin{tabular}{lcccccc} 
        \thickhline
        Benchmark & \multicolumn{3}{c}{MultiPL-E Lua} \\
        Retrieval Method &  MPNet & CodeBERT & BM25  \\
        \hline
        w/o Examples & 49.94 & 49.94 & 49.94  \\
        Random Selection & 60.68 & 60.68 & \underline{60.68} \\ 
        Problem-As-Query Retrieval & 60.93 & 58.76 & 60.37 \\
        CEDAR & \underline{61.61} & \underline{60.99} & 60.25 \\
        RepoCoder & 60.81 & 60.50 & \textbf{63.23} \\
        \rowcolor{gray!10}
        \ours & \textbf{64.10} & \textbf{61.80} & 59.01 \\ 
        \thickhline
    \end{tabular}
    \normalsize
    \caption{Pass@1 scores of GPT-3.5-Turbo-16k with different retrievers.}
    \label{tab:different_retriever}
\end{table}

\clearpage
\section{In-Context Learning Examples}
\label{icl_examples}
We provide concrete examples of prompts used for in-context learning in Tables~\ref{tab:icl_query_exp}, \ref{tab:icl_pseudo_ext}, and \ref{tab:icl_convert_code}. The final prompt is created by appending the target problem to 3-shot examples in the same format as shown in the tables.

\begin{table*}[h]
\begin{tabularx}{\textwidth}{X}
\thickhline
\textbf{Problem Description}\\
def count\_bidirectional(test\_list):\\
$\quad$"""Write a function to count bidirectional tuple pairs.\\
$\quad$"""\\
\hdashline
\textbf{Instruction for Plan Generation}\\
Write a plan for the problem.\\
\hdashline
\textbf{Generated Plan}\\
\end{tabularx}
\renewcommand{\arraystretch}{0.7}
\begin{tabularx}{\textwidth}{X}
\# Define a function to count bidirectional tuple pairs.\\
\# Create a map to store pairs and their reverse occurrences.\\
\# Initialize a counter for bidirectional pairs.\\
\# Iterate over the input list of tuples.\\
$\quad$\# Retrieve the two elements of the tuple.\\
$\quad$\# Create both the original pair and the reverse pair.\\
$\quad$\# Check if the reverse pair exists in the map.\\
$\quad$$\quad$\# If it does, increment the bidirectional pair count and decrement the occurrence of the reverse pair in the map.\\
$\quad$\# If the reverse pair does not exist in the map, increment the occurrence of the original pair in the map.\\
\# Return the count of bidirectional pairs.\\
\thickhline
\end{tabularx}
\caption{The in-context example used for predicting the algorithmic plan from the test problem for query expansion by \ours.
The problem description example is from~\citet{austin2021programsynthesislargelanguage}.}
\label{tab:icl_query_exp}
\end{table*}

\begin{table*}[h]
\begin{tabularx}{\textwidth}{X}
\\
\thickhline
\textbf{Problem Description}\\
def count\_bidirectional(test\_list):\\
$\quad$"""Write a function to count bidirectional tuple pairs.\\
$\quad$"""\\
\hdashline
\textbf{Code}\\
def count\_bidirectional(test\_list):\\
$\quad$pair\_count = \{\}\\
$\quad$bidirectional\_count = 0\\

$\quad$for tup in test\_list:\\
$\quad$$\quad$a, b = tup[0], tup[1]\\
$\quad$$\quad$original = (a, b)\\
$\quad$$\quad$reverse = (b, a)\\

$\quad$$\quad$if reverse in pair\_count and pair\_count[reverse] > 0:\\
$\quad$$\quad$$\quad$bidirectional\_count += 1\\
$\quad$$\quad$$\quad$pair\_count[reverse] -= 1\\
$\quad$$\quad$else:\\
$\quad$$\quad$$\quad$pair\_count[original] = pair\_count.get(original, 0) + 1\\

$\quad$return bidirectional\_count\\
\hdashline
\textbf{Instruction for Plan Generation}\\
Write a plan for the problem.\\
\hdashline
\textbf{Generated Plan}\\
\end{tabularx}
\renewcommand{\arraystretch}{0.7}
\begin{tabularx}{\textwidth}{X}
\# Define a function to count bidirectional tuple pairs.\\
\# Create a map to store pairs and their reverse occurrences.\\
\# Initialize a counter for bidirectional pairs.\\
\# Iterate over the input list of tuples.\\
$\quad$\# Retrieve the two elements of the tuple.\\
$\quad$\# Create both the original pair and the reverse pair.\\
$\quad$\# Check if the reverse pair exists in the map.\\
$\quad$$\quad$\# If it does, increment the bidirectional pair count and decrement the occurrence of the reverse pair in the map.\\
$\quad$\# If the reverse pair does not exist in the map, increment the occurrence of the original pair in the map.\\
\# Return the count of bidirectional pairs.\\
\thickhline
\end{tabularx}
\caption{The in-context example used for converting code in source PL to algorithmic plan by \ours. The problem description example is from~\citet{austin2021programsynthesislargelanguage}.}
\label{tab:icl_pseudo_ext}
\end{table*}

\begin{table*}[h]
\begin{tabularx}{\textwidth}{X}
\\
\thickhline
\textbf{Problem Description}\\
def count\_bidirectional(test\_list):\\
$\quad$"""Write a function to count bidirectional tuple pairs.\\
$\quad$"""\\
\hdashline
\textbf{Code}\\
def count\_bidirectional(test\_list):\\
$\quad$pair\_count = \{\}\\
$\quad$bidirectional\_count = 0\\

$\quad$for tup in test\_list:\\
$\quad$$\quad$a, b = tup[0], tup[1]\\
$\quad$$\quad$original = (a, b)\\
$\quad$$\quad$reverse = (b, a)\\

$\quad$$\quad$if reverse in pair\_count and pair\_count[reverse] > 0:\\
$\quad$$\quad$$\quad$bidirectional\_count += 1\\
$\quad$$\quad$$\quad$pair\_count[reverse] -= 1\\
$\quad$$\quad$else:\\
$\quad$$\quad$$\quad$pair\_count[original] = pair\_count.get(original, 0) + 1\\

$\quad$return bidirectional\_count\\
\hdashline
\textbf{Plan}\\
\end{tabularx}
\renewcommand{\arraystretch}{0.7}
\begin{tabularx}{\textwidth}{X}
\# Define a function to count bidirectional tuple pairs.\\
\# Create a map to store pairs and their reverse occurrences.\\
\# Initialize a counter for bidirectional pairs.\\
\# Iterate over the input list of tuples.\\
$\quad$\# Retrieve the two elements of the tuple.\\
$\quad$\# Create both the original pair and the reverse pair.\\
$\quad$\# Check if the reverse pair exists in the map.\\
$\quad$$\quad$\# If it does, increment the bidirectional pair count and decrement the occurrence of the reverse pair in the map.\\
$\quad$\# If the reverse pair does not exist in the map, increment the occurrence of the original pair in the map.\\
\# Return the count of bidirectional pairs.\\
\hdashline
\textbf{Instruction for Plan Generation}\\
Convert to Lua code for the problem following the plan. \\
\hdashline
\textbf{Converted Code to Target PL}\\
\end{tabularx}
\renewcommand{\arraystretch}{0.7}
\begin{tabularx}{\textwidth}{X}
function count\_bidirectional(test\_list) \\
$\quad$local pair\_count = \{\} \\
$\quad$local bidirectional\_count = 0 \\
 \\
$\quad$for \_, tup in ipairs(test\_list) do \\
$\quad$$\quad$local a, b = tup[1], tup[2] \\
$\quad$$\quad$local original = \{a, b\} \\
$\quad$$\quad$local reverse = \{b, a\} \\

$\quad$$\quad$if pair\_count[reverse] and pair\_count[reverse] > 0 then \\
$\quad$$\quad$$\quad$bidirectional\_count = bidirectional\_count + 1 \\
$\quad$$\quad$$\quad$pair\_count[reverse] = pair\_count[reverse] - 1 \\
$\quad$$\quad$else \\
$\quad$$\quad$$\quad$pair\_count[original] = (pair\_count[original] or 0) + 1 \\
$\quad$$\quad$end \\
$\quad$end \\
 \\
$\quad$return bidirectional\_count \\
end \\
\thickhline
\end{tabularx}
\caption{The in-context example used for converting code in source PL to target PL by \ours. The problem description example is from~\citet{austin2021programsynthesislargelanguage}.}
\label{tab:icl_convert_code}
\end{table*}

\clearpage

\section{Retrieved Example and Generated Code with \ours}
We provide concrete example of the few-shot examples retrieved by PERC and the resultant pseudocode and generated code on HumanEval, in which the target PL is high-resource (python), in Tables~\ref{tab:ours_retrieved_example_high} and \ref{tab:ours_generated_example_high}.
We also provide examples from MultiPL-E-Lua, underrepresented target PL setting, in Tables~\ref{tab:ours_retrieved_example} and \ref{tab:ours_generated_example}.

\begin{table*}[h]
\begin{tabularx}{\textwidth}{X}
\thickhline
\textbf{Problem Description} $t_q$\\
\end{tabularx}
\renewcommand{\arraystretch}{0.7}
\begin{tabularx}{\textwidth}{X}
from typing import List \\
\mbox{} \\
def below\_zero(operations: List[int]) -> bool:\\
$\quad$""" You're given a list of deposit and withdrawal operations on a bank account that starts with\\
$\quad$zero balance. Your task is to detect if at any point the balance of account falls below zero, and\\
$\quad$at that point function should return True. Otherwise it should return False.\\
$\quad$> below\_zero([1, 2, 3])\\
$\quad$False\\
$\quad$> below\_zero([1, 2, -4, 5])\\
$\quad$True\\
$\quad$"""\\
\hdashline
\end{tabularx}
\renewcommand{\arraystretch}{1.0}
\begin{tabularx}{\textwidth}{X}
\textbf{Predicted Pseudocode} $\hat{p}_q$\\
\end{tabularx}
\renewcommand{\arraystretch}{0.7}
\begin{tabularx}{\textwidth}{X}
\# Set the initial balance of the bank account to zero \\
\# Iterate through each operation in the list \\
$\quad$\# Update the balance of the bank account by adding the current operation \\
$\quad$\# Check if the balance is less than zero \\
$\quad$$\quad$\# If it is, return True \\
\\
\# If the balance never falls below zero, return False \\
\hdashline
\end{tabularx}
\renewcommand{\arraystretch}{1.0}
\begin{tabularx}{\textwidth}{X}
\textbf{Retrieved Problem with Pseudocode} $t; p$\\
\end{tabularx}
\renewcommand{\arraystretch}{0.7}
\begin{tabularx}{\textwidth}{X}
Write a function to calculate the sum of the negative numbers of a given list of numbers.\\
\# Initialize the sum of negative numbers to 0\\
\\
\# Iterate through each number in the list\\
$\quad$\# Check if the number is negative\\
$\quad$$\quad$\# If it is negative, add it to the sum\\
\\
\# Return the sum of negative numbers\\
\hdashline
\end{tabularx}
\renewcommand{\arraystretch}{1.0}
\begin{tabularx}{\textwidth}{X}
\textbf{Retrieved Code} $c$\\
\end{tabularx}
\renewcommand{\arraystretch}{0.7}
\begin{tabularx}{\textwidth}{X}
def sum\_negativenum(nums):\\
$\quad$sum\_negativenum = list(filter(lambda nums:nums<0,nums))\\
$\quad$return sum(sum\_negativenum)\\
\thickhline
\end{tabularx}
\caption{The example of pseudocode to extend the query for a target HumanEval problem \#3, along with the problem, pseudocode, code, and converted code in the target PL of the retrieved candidate through \ours.}
\label{tab:ours_retrieved_example_high}
\end{table*}

\begin{table*}[h]
\begin{tabularx}{\textwidth}{X}
\thickhline
\textbf{Problem Description}\\
from typing import List \\
\\
\\
def below\_zero(operations: List[int]) -> bool:\\
$\quad$""" You're given a list of deposit and withdrawal operations on a bank account that starts with\\
$\quad$zero balance. Your task is to detect if at any point the balance of account fallls below zero, and\\
$\quad$at that point function should return True. Otherwise it should return False.\\
$\quad$> below\_zero([1, 2, 3])\\
$\quad$False\\
$\quad$> below\_zero([1, 2, -4, 5])\\
$\quad$True\\
$\quad$"""\\
\hdashline
\textbf{Generated Pseudocode}\\
\end{tabularx}
\renewcommand{\arraystretch}{0.7}
\begin{tabularx}{\textwidth}{X}
\# Initialize a variable "balance" to 0\\
\\
\# Iterate through each operation in the given list\\
$\quad$\# Add the operation to the balance\\
\\
$\quad$\# Check if the balance is less than 0\\
$\quad$$\quad$\# If it is, return True\\
\hdashline
\textbf{Generated Code}\\
def below\_zero(operations: List[int]) -> bool:\\
$\quad$balance = 0\\
$\quad$for operation in operations:\\
$\quad$$\quad$balance += operation\\
$\quad$$\quad$if balance < 0:\\
$\quad$$\quad$$\quad$return True\\
$\quad$return False\\
\thickhline
\end{tabularx}
\caption{An example of pseudocode and code generated using few-shot prompting with \ours for HumanEval problem \#3, based on examples from Table~\ref{tab:ours_retrieved_example_high}.}
\label{tab:ours_generated_example_high}
\end{table*}

\begin{table*}[h]
\begin{tabularx}{\textwidth}{X}
\thickhline
\textbf{Problem Description} $t_q$\\
-- You're given a table of deposit and withdrawal operations on a bank account that starts with \\
-- zero balance. Your task is to detect if at any point the balance of account falls below zero, and \\
-- at that point function should return true. Otherwise it should return false. \\
-- > below\_zero({1, 2, 3}) \\
-- false \\
-- > below\_zero({1, 2, -4, 5}) \\
-- true \\
local function below\_zero(operations) \\
\hdashline
\textbf{Predicted Pseudocode} $\hat{p}_q$\\
\end{tabularx}
\renewcommand{\arraystretch}{0.7}
\begin{tabularx}{\textwidth}{X}
\# Set the initial balance of the bank account to zero \\
\# Iterate through each operation in the list \\
$\quad$\# Update the balance of the bank account by adding the current operation \\
$\quad$\# Check if the balance is less than zero \\
$\quad$$\quad$\# If it is, return True \\
\\
\# If the balance never falls below zero, return False \\
\hdashline
\end{tabularx}
\renewcommand{\arraystretch}{1.0}
\begin{tabularx}{\textwidth}{X}
\textbf{Retrieved Problem with Pseudocode} $t; p$\\
\end{tabularx}
\renewcommand{\arraystretch}{0.7}
\begin{tabularx}{\textwidth}{X}
Write a python function to return the negative numbers in a list.\\
\\
\# Initialize an empty list to store the negative numbers\\
\# Iterate through each element in the input list\\
$\quad$\# Check if the current element is less than 0\\
$\quad$$\quad$\# If the element is negative, append it to the list of negative numbers\\
\\
\# Return the list of negative numbers \\
\hdashline
\end{tabularx}
\renewcommand{\arraystretch}{1.0}
\begin{tabularx}{\textwidth}{X}
\textbf{Retrieved Code} $c$\\
\end{tabularx}
\renewcommand{\arraystretch}{0.7}
\begin{tabularx}{\textwidth}{X}
def neg\_nos(list1):\\
$\quad$out = []\\
$\quad$for num in list1: \\
$\quad$$\quad$if num < 0: \\
$\quad$$\quad$$\quad$out.append(num)\\
$\quad$return out\\
\hdashline
\end{tabularx}
\renewcommand{\arraystretch}{1.0}
\begin{tabularx}{\textwidth}{X}
\textbf{Converted Code in Target PL} $\hat{c}$ \\
\end{tabularx}
\renewcommand{\arraystretch}{0.7}
\begin{tabularx}{\textwidth}{X}
function getNegativeNumbers(numbers)\\
$\quad$local negativeNumbers = \{\}\\
\\
$\quad$for \_, num in ipairs(numbers) do\\
$\quad$$\quad$if num < 0 then\\
$\quad$$\quad$$\quad$table.insert(negativeNumbers, num)\\
$\quad$$\quad$end\\
$\quad$end\\
\\
$\quad$return negativeNumbers\\
end\\
\thickhline
\end{tabularx}
\caption{The example of pseudocode to extend the query for a target HumanEval-Lua problem \#3, along with the problem, pseudocode, code, and converted code in the target PL of the retrieved candidate through \ours.}
\label{tab:ours_retrieved_example}
\end{table*}

\begin{table*}[h]
\begin{tabularx}{\textwidth}{X}
\thickhline
\textbf{Problem Description}\\
-- You're given a table of deposit and withdrawal operations on a bank account that starts with \\
-- zero balance. Your task is to detect if at any point the balance of account fallls below zero, and \\
-- at that point function should return true. Otherwise it should return false. \\
-- > below\_zero({1, 2, 3}) \\
-- false \\
-- > below\_zero({1, 2, -4, 5}) \\
-- true \\
local function below\_zero(operations) \\
\hdashline
\textbf{Generated Pseudocode}\\
\end{tabularx}
\renewcommand{\arraystretch}{0.7}
\begin{tabularx}{\textwidth}{X}
\# Return the count of bidirectional pairs.\\
\# Initialize the balance of the account to 0 \\
\# Iterate through each operation in the list \\
$\quad$\# Add the operation to the balance \\
$\quad$\# Check if the balance is less than 0 \\
$\quad$$\quad$\# If it is, return true \\
\# If the loop completes without returning true, return false\\
\hdashline
\textbf{Generated Code}\\
function below\_zero(operations) \\
$\quad$local balance = 0 \\
\\
$\quad$for \_, operation in ipairs(operations) do \\
$\quad$$\quad$balance = balance + operation \\
\\
$\quad$$\quad$if balance < 0 then \\
$\quad$$\quad$$\quad$return true \\
$\quad$$\quad$end \\
$\quad$end \\
\\
$\quad$return false \\
end \\
\thickhline
\end{tabularx}
\caption{An example of pseudocode and code generated using few-shot prompting for the target HumanEval-Lua problem \#3, based on PERC-retrieved examples from Table~\ref{tab:ours_retrieved_example}.}
\label{tab:ours_generated_example}
\end{table*}

\end{document}